\documentclass[preprint,1p]{elsarticle}

\usepackage{amssymb}
\usepackage{lineno}
\usepackage{hyperref}
\usepackage[percent]{overpic}
\usepackage{enumitem}

\newcommand*{\NAZ}{\rho_A(0)}
 
\newcommand{\ssi}{\sigma_{Acc}}
\newcommand{\kap}{k_{cap}}
 
\newcommand{\nq[1]}{\cdot 10^{#1} \; n_{eq}/cm^2 } 
\newcommand{\VNA}{$\rho_{Ao} = 2.5 \cdot 10^{16} \; n/cm^3 \;$}
\newcommand{\VSC}{$\ssi = 76$ mb }
\newcommand{\vglc}{V^C_{GL}} 
\newcommand{\vglr}{V^R_{GL}} 
\newcommand{\vgl}{V_{GL}} 
\newcommand{\er}{\pm \; 1.0}
\newcommand{\err}{\pm \; 1.5}

\begin{document}

\begin{frontmatter}


\title{Radiation resistant  LGAD design}

\author[1]{M.~Ferrero}
\author[1,2]{R.~Arcidiacono}
\author[3,5]{M.~Barozzi}
\author[3,5]{M.~Boscardin}
\author[1]{N.~Cartiglia\corref{cor}}\ead{cartiglia@to.infn.it}
\author[4,5]{G.F.~Dalla Betta}
\author[7]{Z.~Galloway}
\author[1]{M.~Mandurrino}
\author[7]{S.~Mazza}
\author[3,5]{G.~Paternoster}
\author[3,5]{F.~Ficorella}
\author[4,5]{L.~Pancheri}
\author[7]{H-F W.~Sadrozinski}
\author[1,6]{F. Siviero}
\author[1,6]{V.~Sola}
\author[1]{A.~Staiano}
\author[7]{A.~Seiden}
\author[1,6]{M. Tornago}
\author[7]{Y.~ Zhao}

\address[1]{INFN, Torino, Italy}
\address[2]{Universit\`a del Piemonte Orientale, Italy}
\address[3]{Fondazione Bruno Kessler, Trento, Italy}
\address[4]{Universit\`a di Trento, Trento, Italy}
\address[5]{TIFPA-INFN, via Sommarive 18, 38123, Povo (TN), Italy}
\address[6]{Universit\`a di Torino, Torino, Italy}
\address[7]{SCIPP, University of California Santa Cruz, CA, USA}

\cortext[cor]{Corresponding author}

\begin{abstract}
In this paper,  we report on the radiation resistance of 50-micron thick Low Gain Avalanche Diodes (LGAD)  manufactured at the Fondazione Bruno Kessler (FBK) employing  different  dopings in the gain layer.  LGADs with a gain layer made of  Boron, Boron low-diffusion, Gallium, Carbonated Boron and Carbonated Gallium  have been designed and successfully produced at FBK. These sensors have been  exposed to neutron  fluences up to $\phi_n \sim 3 \cdot 10^{16}\; n/cm^2$ and to proton fluences up to $\phi_p \sim 9\cdot10^{15}\; p/cm^2$ to test their radiation resistance. The experimental results show that Gallium-doped LGAD are more heavily affected  by the initial acceptor removal mechanism than those doped with Boron, while the addition of Carbon reduces this effect both for Gallium and Boron doping.  The Boron low-diffusion gain layer shows a higher radiation resistance than that of standard Boron implant, indicating a dependence of  the initial acceptor removal mechanism upon the implant density. 
\end{abstract}


\end{frontmatter}



The LGAD design evolves the standard silicon sensors design by incorporating low, controlled gain~\cite{LGAD1} in the signal formation mechanism. The overarching idea is to manufacture silicon detectors with signals large enough to assure excellent timing performance while  maintaining almost unchanged levels of noise \cite{ROPP}. 

\vspace{2mm}

\begin{figure}[htb]
\begin{center}
\includegraphics[width=0.6\textwidth]{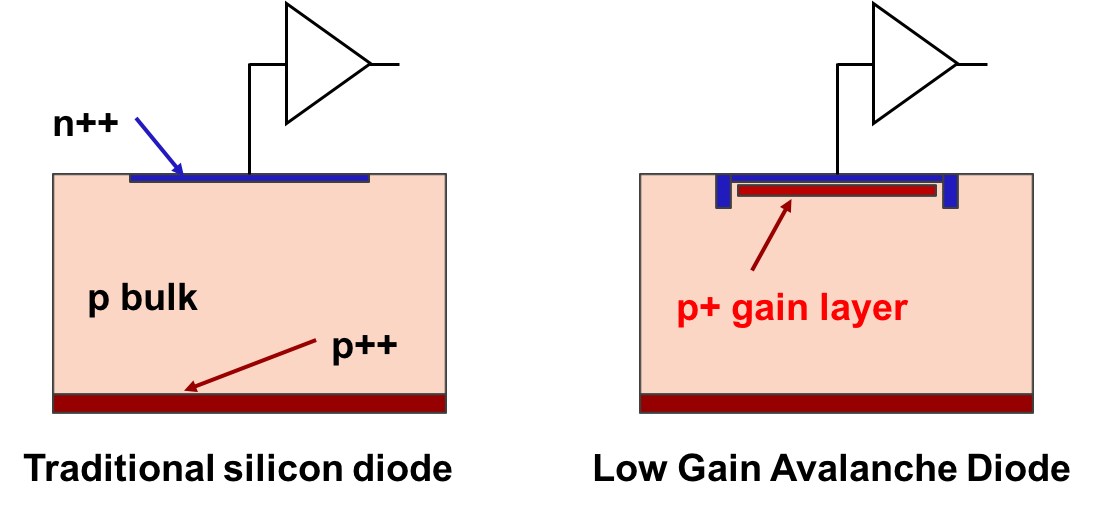}
\caption{Schematic of a traditional silicon diode (left) and of a Low-Gain Avalanche Diode (right). The additional $p^+$ layer underneath the $n^{++}$ electrode creates, when depleted, a large electric field that generates charge multiplications.}
\label{fig:LGAD}
\end{center}
\end{figure}

Charge multiplication in silicon sensors happens when the charge carriers (electrons and holes) are in electric fields of the order of $ E \sim 300$ kV/cm \cite{ImpIon}. Under this condition, the electrons (and to less extent the holes) acquire sufficient kinetic energy to generate additional e/h pairs by impact ionization. Field values of $\sim$300 kV/cm  can be obtained by implanting an appropriate acceptor (or donor) charge density $\rho_A$   (of the order $\rho_A \sim 10^{16}/cm^3$)  that, when depleted,  locally generates very high fields.  For this reason, an additional doping layer has been added at the $n-p$ junction in the LGAD design, Figure~\ref{fig:LGAD}.

\section{Initial acceptor removal in LGAD sensors}
It has been shown in previous studies \cite{GK15, Galloway} that neutrons and charged hadrons irradiations  reduce the value of gain  in LGADs. This effect is due to the \emph{initial acceptor removal} mechanism that progressively  deactivates the  acceptors forming the gain layer.  The effects of initial acceptor removal on the silicon sensor bulk  has been  first measured in  standard Boron-doped silicon sensors  more than 20 years ago~\cite{Unno}. Concurrently with the initial acceptor removal mechanism, irradiation causes also the creation of acceptor-like defects due to the creation of deep traps. The combined effects are described by  equation (\ref{eq:ac})~\cite{ROPP, MM-IEEE17}

\begin{equation}
\label{eq:ac}
\rho_A(\phi) = g_{eff} \phi + \NAZ e^{-c\phi}, 
\end{equation}

where $g_{eff}$ = 0.02  [cm$^{-1}$] (see for example chapter 5 of \cite{Balbuena:1291631}),  $\phi$ the irradiation fluence   [ cm$^{-2}$],  $\NAZ$  $(\rho_{A}(\phi))$ the initial (after a fluence $\phi$) acceptor density [cm$^{-3}$], and $c$ [cm$^2$] is a constant  that depends on the initial acceptor concentration $\NAZ$ and on the type of irradiation. 
The first term of equation (\ref{eq:ac}) accounts for acceptor creation by deep traps  while the second term for the initial acceptor removal mechanism. The factor  $c$ can be rewritten as $\phi_o = 1/c$, making more apparent its meaning: $\phi_o$ is the fluence needed to reduce the initial doping density $\NAZ$ to 1/e of its initial value.

\vspace{2mm}

The microscopic origin of the acceptor removal mechanism has not been fully understood, however, it is plausible  that the  progressive inactivation of the Boron atoms with irradiation happens  via the formation of  ion-acceptor complexes.  In this model, the active (substitutionals) doping elements are removed from their lattice sites due to a 2-step process: (i) the  radiation produces interstitial Si atoms that subsequently (ii) inactivate the doping elements via kick-out reactions (Watkins mechanism~\cite{Wa}) that produce ion-acceptor complexes (interstitials)~\cite{Wu}.  

\begin{figure}[htb]
\begin{center}
\includegraphics[width=0.6\textwidth]{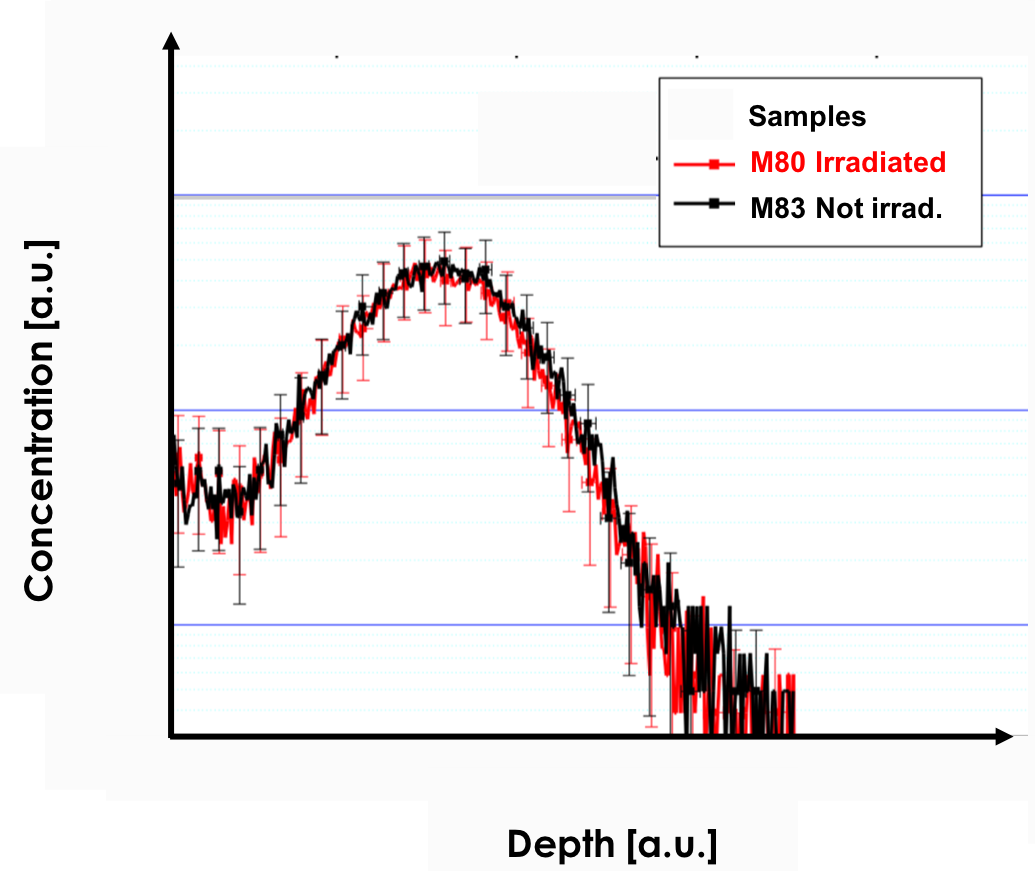}
\caption{Density of Boron atoms forming the gain layer in a new  (M83)  and a heavily irradiated (M80, irradiated to $1\nq[16]$)  LGAD. Even though the gain layer of the M80 sensor is almost completely deactivated, M83 and M80 have the same gain layer doping profile (the plot has log-y and lin-x axis).} 
\label{fig:sims}
\end{center}
\end{figure}

\vspace{2mm} 
Secondary Ion Mass Spectrometer (SIMS) measurements support this view: Figure~\ref{fig:sims} shows the densities of Boron atoms  forming the gain layer as a function of depth in a new  (M83)  and a heavily irradiated (M80, irradiated to $\phi \sim 1\nq[16]$)  LGAD where the gain layer has completely disappeared. The SIMS were performed  in the central area of 1 mm$^2$ LGADs.   Remarkably, the SIMS results are almost identical: the decrease of the active gain layer doping in irradiated sensors does not correspond to a disappearance of the Boron atoms, only  to their inactivation. The SIMS were performed  in the central area of 1 mm$^2$ LGADs 

\subsection{A parametrization of the acceptor removal mechanism}

In a  simple model of acceptor removal,  the number of  initial acceptor atoms deactivated by radiation is given by the product of the fluence  $\phi_o$ times the silicon atomic density  $\rho_{Si} $ times the cross section for an impinging particle to  deactivate an acceptor $\ssi$:

\begin{eqnarray}
\label{eq:ac3}
(1 - 1/e)\NAZ   &=& \phi_o  \cdot \rho_{Si}  \cdot\ssi ,  \\
\NAZ  &=& \frac{1}{0.63} \phi_o \cdot  \rho_{Si}  \cdot \ssi,
\end{eqnarray}

where $\rho_{Si}  = 5\cdot 10^{22} \; cm^{-3}$. 
Following the two-step model outlined above, the expression of $\ssi$ can be written as  the product of the cross section between radiation and Silicon  ($\sigma_{Si}$) times the number of interstitials generated in the scattering ($N_{Int}$)  times the probability of capturing an acceptor ($\kap$): 
\begin{equation}
\label{eq:ssi}
\ssi = \kap \cdot N_{Int} \cdot \sigma_{Si}.
\end{equation}

 Note that the presence of impurities (Carbon, Oxigen,...) influence the value of $\kap$ as they might intercept the interstitial atoms before they reach the acceptors. 

\begin{figure}[htb]

\begin{center}
\includegraphics[width=0.8\textwidth]{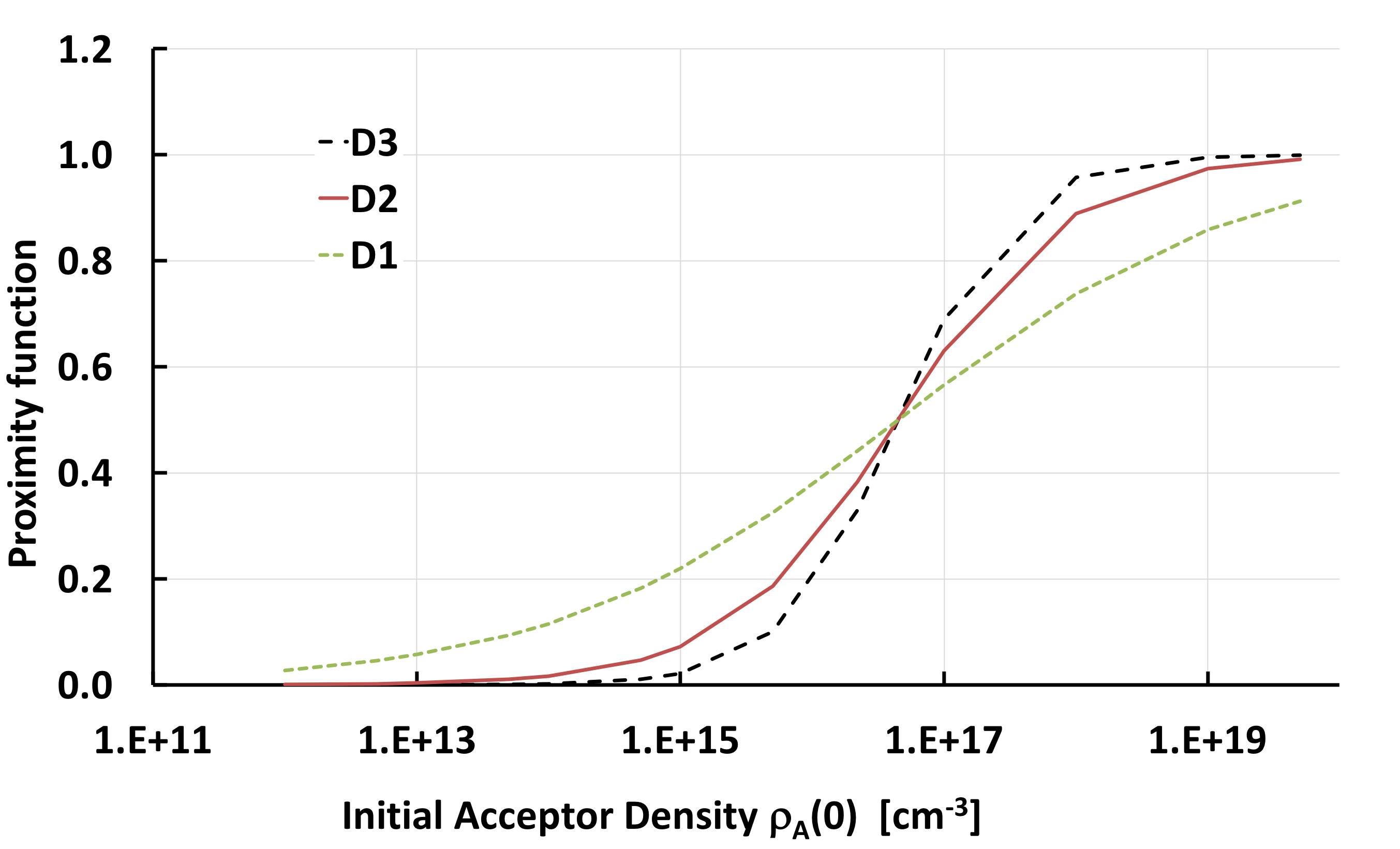}
\caption{Proximity functions D1, D2, and D3. The value \VNA has been used in this plot. }
\label{fig:den}
\end{center}
\end{figure}

\vspace{2mm}
 Equation~(\ref{eq:ac3}) assumes  that each interstitial atom created by radiation is in the proximity of acceptors, however this might not be the case at low acceptor density. For this reason, a proximity  function $D$ needs to be included in equation~(\ref{eq:ac3}): this function describes the probability that an interstitial atom is in the vicinity of an acceptor atom. The analytic form of $D$ is not unique, any smooth function that  goes to 0 at low acceptor density and to 1 at large density is acceptable,  for example:

\begin{equation}
\label{eq:ac4}
Dn = \frac{1}{1+(\frac{\rho_{Ao}}{\NAZ})^{n/3}}, 
\end{equation}
where $\rho_{Ao}$ is a fit parameter indicating the acceptor density at which an interstitial state has a probability of 0.5 of being in the vicinity of an acceptor and $n$ is an exponent that needs to be determined experimentally.  
Figure~\ref{fig:den} shows the values of D1, D2 and D3 ($n$ = 1, 2 or 3) with \VNA.

\begin{figure}[htb]
\begin{center}
\includegraphics[width=0.9\textwidth]{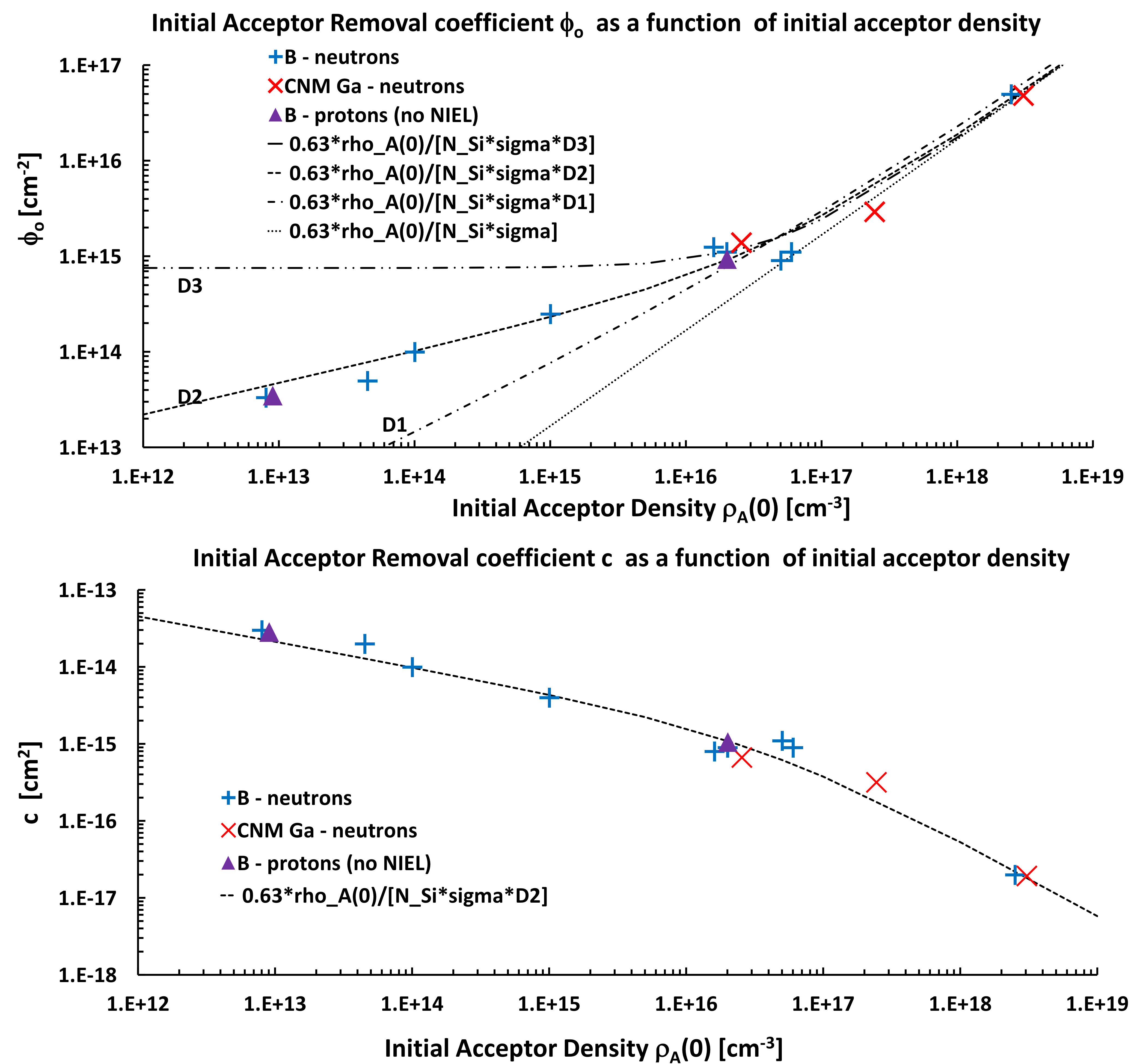}
\caption{The two plots show the parametrization of  $c$ (bottom) and $\phi_o$ (top) from equation~(\ref{eq:ac5})  together with experimental points  as a function of the initial acceptor density. The top plot also shows the parametrization of  equation~(\ref{eq:ac5}) with and without the effect of the proximity functions. The best agreement data - parametrization is obtained with the D2 proximity  function.}
\label{fig:coef}
\end{center}
\end{figure}

 Combining equations~(\ref{eq:ac3}) and~(\ref{eq:ac4}),  the expression linking  the fluence $\phi_o$ to the number of deactivated acceptors is: 


\begin{eqnarray}
\label{eq:ac5}
\phi_o  \cdot \rho_{Si}  \cdot \ssi  \frac{1}{1+(\frac{\rho_{Ao}}{\NAZ})^{n/3}}  &=& 0.63 \NAZ,  \\
\phi_o &=& 0.63 \frac{ \NAZ}{\rho_{Si}  \cdot \ssi} (1+(\frac{\rho_{Ao}}{\NAZ})^{n/3}), 
\end{eqnarray}

where  $\ssi$ and $\rho_{Ao}$ are fit parameters.  Analytic expressions of $Dn$ using a linear  (D1), a surface (D2) and a volumetric (D3) proximity  function were tried, finding the best agreement between models and data with $n$ = 2, \VSC, and \VNA.  The $n = 2$ result indicates that the clusters have a cylindrical shape since spherical shape would have yield to $n = 3$.  Using these numbers, the parameterizations of  equation~(\ref{eq:ac5}) without  the  proximity  function and with each of the three  functions (D1, D2, and D3) are superposed in Figure~\ref{fig:coef} to experimental points. The experimental points of B - neutrons (Boron gain layer irradiated with reactor neutrons)  are taken from ~\cite{1748-0221-10-07-P07006, GK17, HSTD11}, the B - protons  (Boron gain layer irradiated with 800 MeV/c protons) from ~\cite{1748-0221-10-07-P07006, HSTD11} while  Ga - neutrons (Gallium gain layer irradiated with reactor neutrons) from  ~\cite{TREDI2017, KRAMBERGER201853}. 

\vspace{2mm}

The effect of the proximity  function is important at low initial acceptor density, where the overlap probability between interstitial states and acceptors is small and therefore a higher fluence is needed to have initial acceptor removal.  It is important to stress that the acceptor removal rate might differ upon the irradiation type (pions, protons, neutrons), the irradiation energy, and  the acceptor element (Boron or Gallium), however, for lack of statistics, Figure~\ref{fig:coef}  shows a single common fit. 

\vspace{2mm}
The inverse of the density \VNA provides a rough indication of the average volume of   each cluster of defects created by a particle: \VNA yields to a cluster size of $ d = 460$ \AA, which is compatible with the current estimates ~\cite{Meroli,HUHTINEN2002194}.

\vspace{2mm}

\begin{figure}[htb]
\begin{center}
\includegraphics[width=0.9\textwidth]{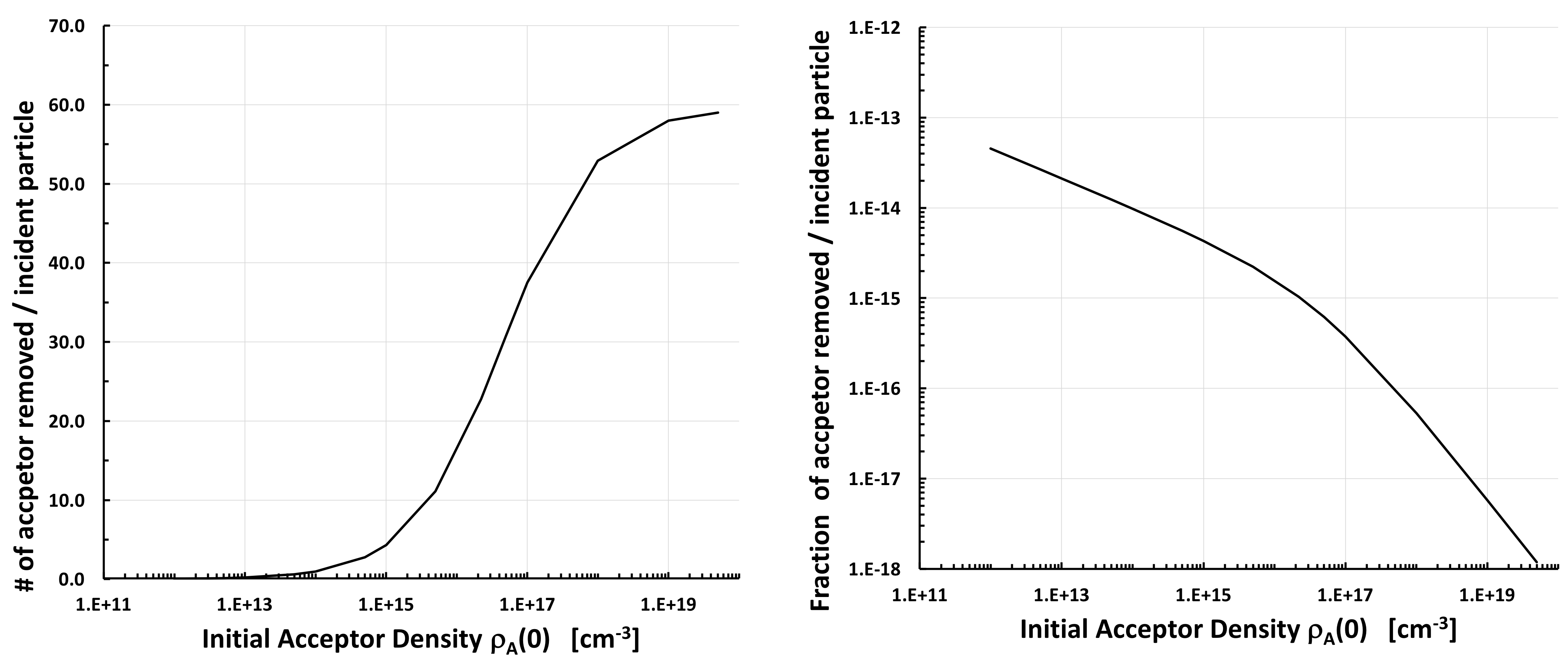}
\caption{The left plot shows the number of removed acceptor atoms per incidente particle: at the highest acceptor density $\sim$ 60 acceptors are removed per incidente particle. The right plot shows instead the fraction of acceptors removed per incident particle demonstrating that the importance of the acceptor removal mechanism is larger at low $\NAZ$ values.}
\label{fig:nr}
\end{center}
\end{figure}

Using the D2 parametrization, the absolute and relative effect of radiation on the initial acceptor density can be studied. The left plot of Figure~\ref{fig:nr}  reports the number of removed acceptors per incident particle as a function of $\NAZ$: it varies from 1 at $\NAZ = 10^{13} \; [cm^{-3}]$ to $\sim$ 60 at $\NAZ = 10^{19} \; [cm^{-3}]$.  Even though the  number of removed acceptors increases with $\NAZ$, the fraction of removed acceptor is strongly decreasing as a function of $\NAZ$ (Figure~\ref{fig:nr}, right plot) demonstrating that high initial acceptor densities are less affected by radiation.

\vspace{2mm}

From the asymptotic behavior of the left plot of Figure~\ref{fig:nr} we can measure the product  $N_{Int}*\kap$, and combining this value with the value of \VSC,  we can calculate $\sigma_{Si}$:

\begin{eqnarray}
\kap \cdot N_{Int} &\sim& 60, \\
\sigma_{Si}  = \frac{\ssi}{\kap \cdot N_{Int}} &\sim& 1.3 \; mb.
\end{eqnarray}

Both numbers are consistent with the results shown in \cite{HUHTINEN2002194} for 1 MeV neutron on Silicon: $\sigma_{Si} \sim$ 4 mb and $N_{Int} \sim 200-300$.

\vspace{2mm}
Finally, using the terms described above,  the expression of the $c$ coefficient  can be written as:

\begin{equation}
\label{eq:ccoef}
c  = \kap \cdot \frac{\rho_{Si}  \cdot N_{Int} \cdot\sigma_{Si}}{0.63 \NAZ}\frac{1}{1+(\frac{\rho_{Ao}}{\NAZ})^{2/3}},
\end{equation}
where the capture coefficient $\kap$ depends upon the doping used for the gain layer and the presence of additional impurities such as Carbon or Oxygen. 

\vspace{2mm}

\begin{figure}[htb]
\begin{center}
\includegraphics[width=0.8\textwidth]{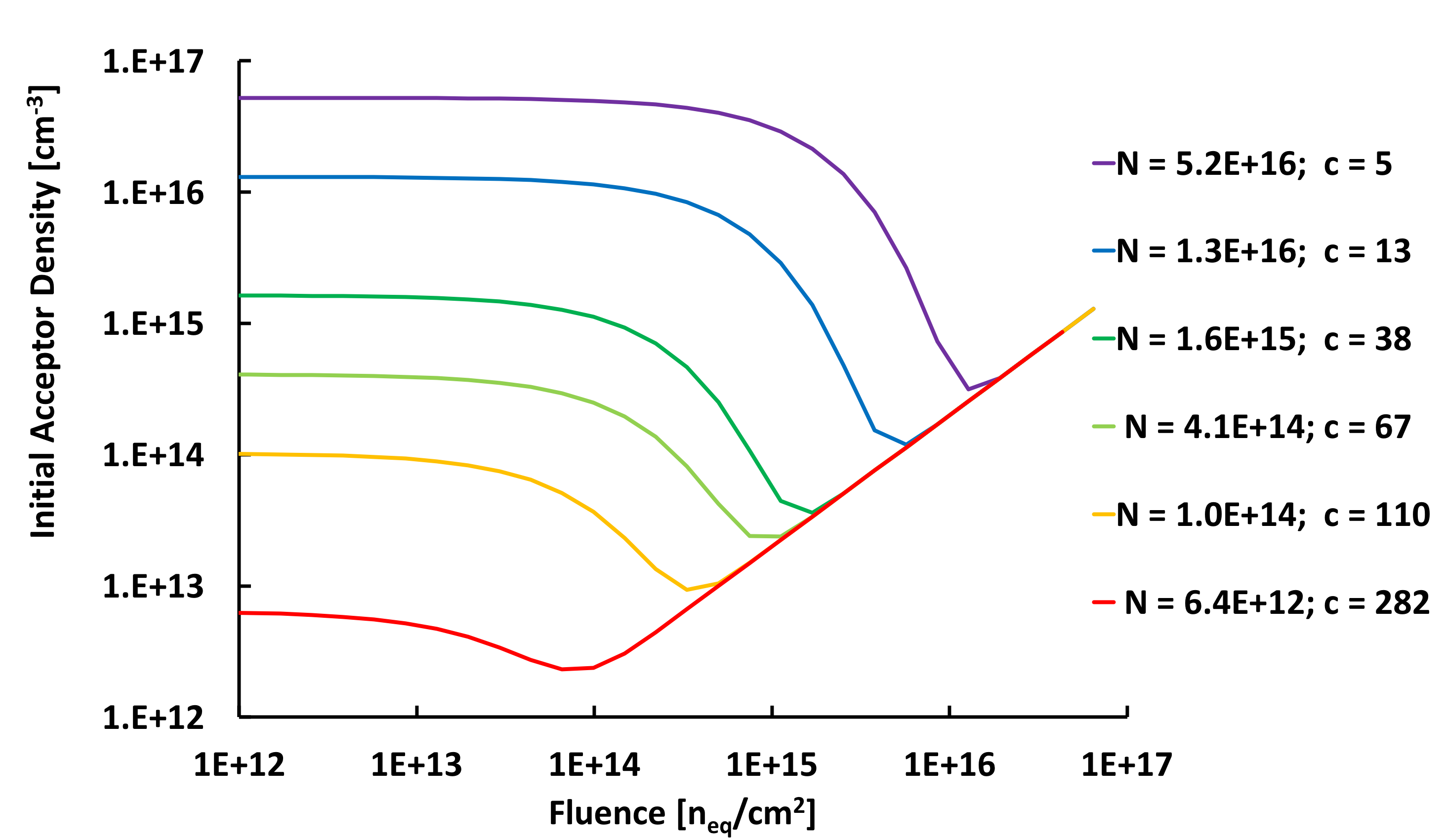}
\caption{Evolution of acceptor density as a function of neutron fluence  for different initial acceptor densities. The lowest acceptor concentration, $\rho_A = 6 \cdot 10^{12}  N/cm^3$, corresponds to the bulk of a high resistivity PiN sensor. The curves have been obtained with a value of $g_{eff} = 0.02$.  The legend reports for each curve the initial acceptor density (in unit of [$N/cm^3$]) and the value of $c$ (indicated in the legend in unit of [$10^{-16} cm^2$]) as obtained from the parametrization D2 shown in Figure~\ref{fig:coef}.}
\label{fig:ar}
\end{center}
\end{figure}

Acceptor creation and initial acceptor removal mechanisms described by equation (\ref{eq:ac})  happen concurrently in the multiplication layer as well as in the bulk.  The evolutions of several initial doping densities  as a function of neutron fluence are shown schematically in Figure~\ref{fig:ar}: the initial Boron doping is removed as the fluence increases and in the meantime new acceptor-like states are created.  At sufficiently high  values of fluence, all initial doping values converge  on the doping density of the high resistivity PiN diodes, indicating a complete disappearance of the initial acceptor density.  

\clearpage

\section{Production of LGAD with different gain layer doping}
Three hypotheses have been put forward for the design of more radiation hard LGADs:  (i) it has been reported  in  \cite{KRAMBERGER201853, KHAN2003271}  that Gallium  might be less prone than Boron to the Watkins mechanism, (ii) the presence of Carbon  atoms might slow down the acceptor removal mechanism by producing ion-carbon complexes instead of  ion-acceptor complexes, and   (iii) a narrower doping layer with higher initial doping should be less prone to the acceptor removal mechanism than a wider doping layer with a lower initial doping. 

\vspace{2mm}

To test these hypotheses, 50-micron thick LGAD sensors with 5
different gain layer configurations have  been manufactured at the
Fondazione Bruno Kessler \footnote{FBK, Fondazione Bruno Kessler,  Trento, Italy} : (i) Boron (B),  (ii) Boron low-diffusion (B LD), (iii) Gallium (Ga), (iv) carbonated Boron (B+C), and (v) carbonated Gallium (Ga+C). This production is called \emph{UFSD2}.  It is important to note that carbon enrichment has been done uniquely in the volume of the gain layer to avoid a sharp increase of the leakage current. Details on the production have been presented in ~\cite{Pater-RD50},  a short summary of the UFSD2 production is shown in Table~\ref{table:UFSD2}: 18 6-inch wafers were processed, 10 with a B-doped and 8 with a Ga-doped gain layer. The B-doped gain layer wafers  W3-10 have 3 splits dose, in 2\% steps, while the Ga-doped gain layer  wafers W11-19 have also 3 splits of dose, however in 4\% steps. Two splits of B-doped and  one of the Ga-doped gain layers have been  co-implanted with Carbon, with two different doses of Carbon. Two wafers with  a B-doped gain layer (W1,2) were exposed to a reduced thermal load during production to minimize the diffusion of Boron (Boron low-diffusion). The Ga-doped wafers, given the higher diffusivity of Gallium, were also exposed to a reduced thermal load, however, the width of the resulting  Gallium implant is nevertheless wider even than that of  the B-doped gain layer with a high thermal load.

\begin{table}[h]
\begin{center}
\begin{tabular}{|c|c|c|c|c|c|}
\hline
Wafer $\#$ &	Dopant	& Gain Dose &	Carbon & Diffusion &  irradiation \\ \hline \hline
{1} &	Boron &	0.98 & 	&  Low  & n \\
	2 &	Boron	 &1.00 &		 &	Low   &\\
	3 &	Boron	 &1.00 &		 &	High  & p \\
	4 &	Boron	 &1.00 &	Low 	 &	High &\\
	5 &	Boron	 &1.00 &	High 	 &	High &  \\
	{6} &	Boron	 &1.02 &	Low	  &	High  & p, n\\
	7 &	Boron	 &1.02 &	High	  &	High  &\\
	{8} &	Boron	 &1.02 &		 &	High  &  n \\
	9 &	Boron	 &1.02 &		 &	High  &\\
	10 &	Boron	 &1.04 &		 &	High  &\\ \hline
	11 &	Gallium	 &1.00 &		 &	Low  &\\
	12 &	Gallium	 &1.00 &		 &	Low  &	\\	
	13 &	Gallium	 &1.04 &		 &	Low  &\\
	{14} &	Gallium	 &1.04 &		 &	Low & p, n\\
	{15} &	Gallium	 &1.04 &	Low	 &	Low & p, n\\
	16 &	Gallium	 &1.04 &	High	  &	Low  &\\
	18 &	Gallium	 &1.08 &		 &	Low &\\
	19 &	Gallium	 &1.08 &		 &	Low  &\\ \hline \hline
\end{tabular}
\caption{Summary of the doping splits in the UFSD2 production. The last column reports the irradiation  campaign (p = protons, n = neutrons). }
\label{table:UFSD2}
\end{center}
\end{table}

\vspace{2mm}

\begin{figure}[htb]
\begin{center}
\includegraphics[width=0.5\textwidth]{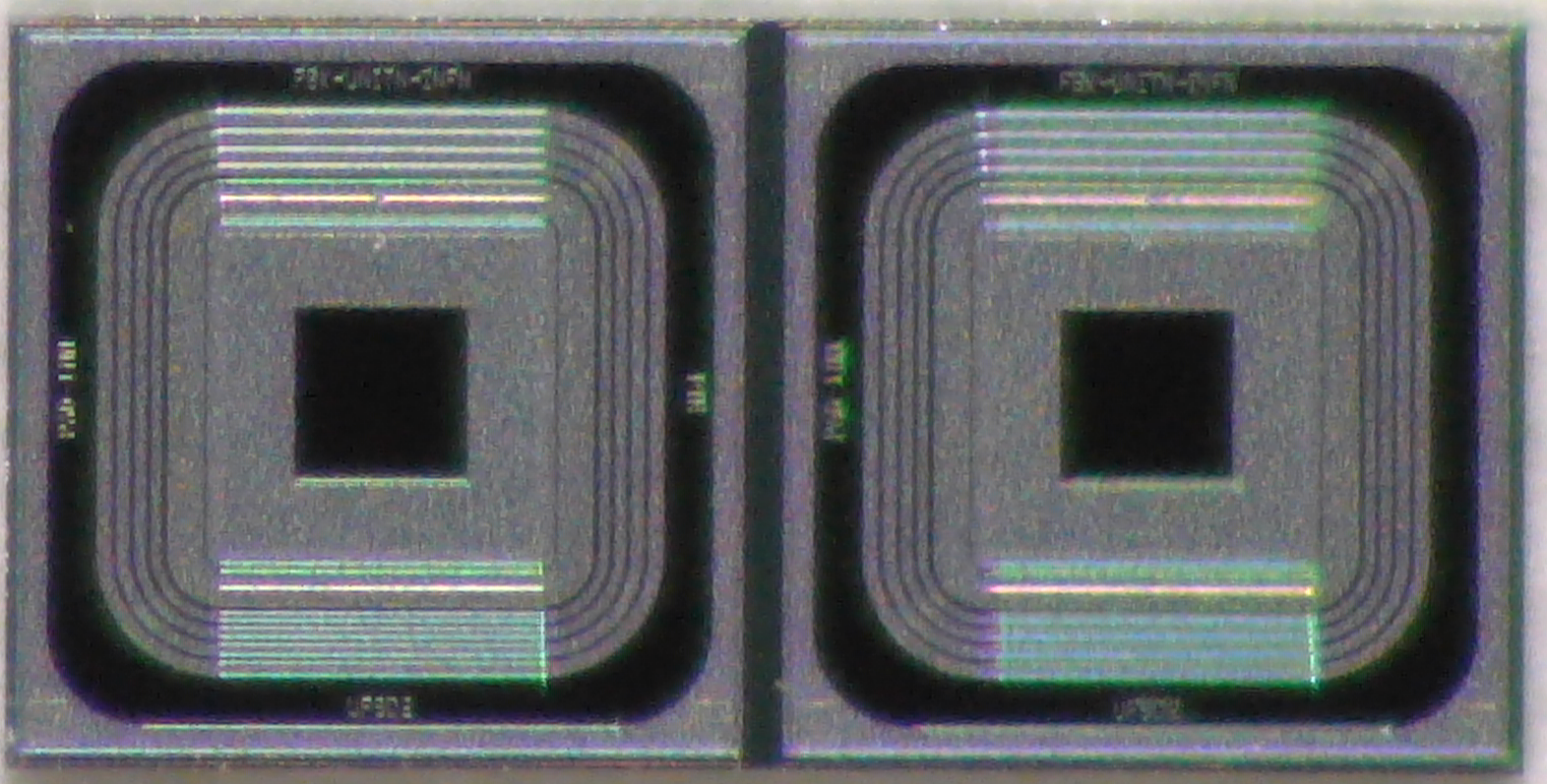}
\caption{Example of a pair PiN-LGAD with 4 guard-rings manufactured by FBK  used in the analysis presented in this work. Each sensor is 1x1 mm$^2$ and 50-micron thick.}
\label{fig:pad}
\end{center}
\end{figure}

UFSD2 layout comprises of many hundreds of devices, from $1\times1 \; mm^2$ single diodes to large arrays of pads and strips~\cite{Pater-RD50}. For this irradiation campaign, pairs of  $1 \times 1  \; mm^2$ PiN - LGAD diodes were used, as shown in  Figure~\ref{fig:pad}. Combined PiN-LGAD irradiation is a very useful tool in assessing the evolution of the LGAD behavior with fluence, as at each irradiation step the PiN  diodes are used as a reference. 

\subsection{Properties of LGAD with different gain layer doping}
\label{sub:lgad}
\begin{figure}[htb]
\begin{center}
\includegraphics[width=0.8\textwidth]{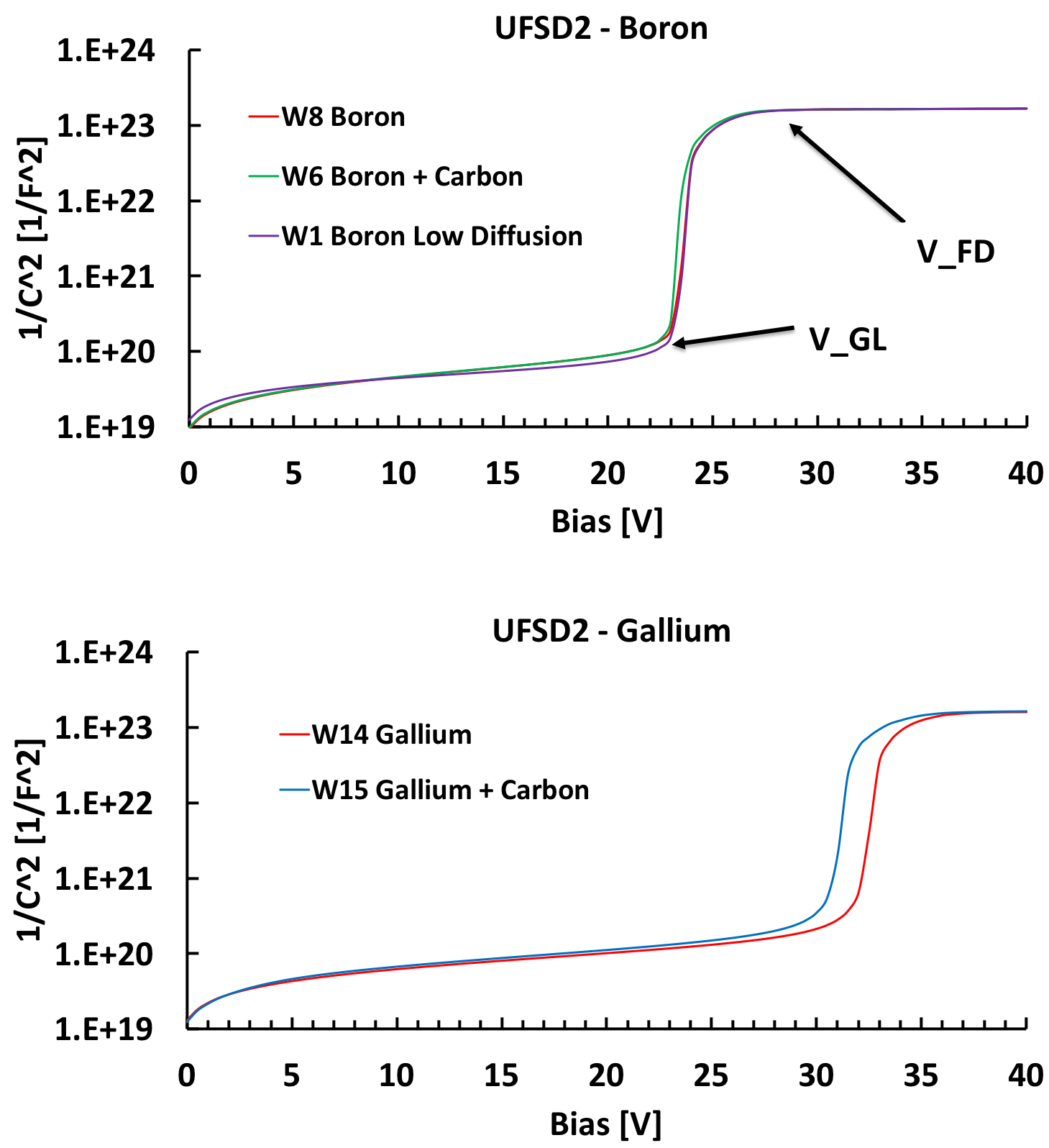}
\caption{Average 1/C$^2$-V curves for each of the wafer used in the irradiation campaign. The lables on the left plot  indicate the points where the gain layer and the bulk deplete. Each curve is the average of 40 diodes. }
\label{fig:IV_CV}
\end{center}
\end{figure}

Figure~\ref{fig:IV_CV} shows on the top pane representative 1/C$^2$-V curves for B and B+C  doped gain  layers LGADs  while on the bottom those of Ga and Ga+C doped gain layers.
The voltage necessary to deplete the gain layer, $V_{GL}$, is proportional to the average active doping $\rho_A$ in the gain layer: 

\begin{equation}
\label{eq:gl}
 V_{GL}=  \frac{q\rho_A}{2\epsilon} w^2
\end{equation}

where  $w$ is the thickness of the gain layer, normally $\sim 1 \mu m$, and $q$ the electron electric charge. Assuming a constant value of $w$,  $V_{GL}$ is directly proportional to $\rho_A$. In the  1/$C^2$-V curves, $\vgl$ can be recognized as the point where the 1/C$^2$-V curve starts a sharp increase, while the voltage of the diode full depletion, $V_{FD}$, is where the 1/C$^2$ becomes constant.  The voltage difference between  $V_{FD}$ and $V_{GL}$, $\Delta V_{Bulk} = V_{FD} - V_{GL}$, is proportional to the doping of the sensor bulk. For non irradiated sensors, as those shown in Figure ~\ref{fig:IV_CV},   $\Delta V_{Bulk} $ is of the order of a few volts indicating a doping of $\rho_{Bulk} \sim 2-3 \cdot 10^{12}$  atoms/cm$^3$. We indicate $\vgl$ measured with the 1/$C^2$-V curves with the symbol $\vglc$.
 
 It is visible in the plot that  Carbon implantation reduces the activated fraction of Gallium,  while the Carbon effects on Boron is minimal: $\vgl$ is on average 0.3V smaller for  B+C LGADs with respect to that of B LGADs. A discussion of the effects of Carbon co-implantation can be found in \cite{shimizu}. 

\begin{figure}[htb]
\begin{center}
\includegraphics[width=0.9\textwidth]{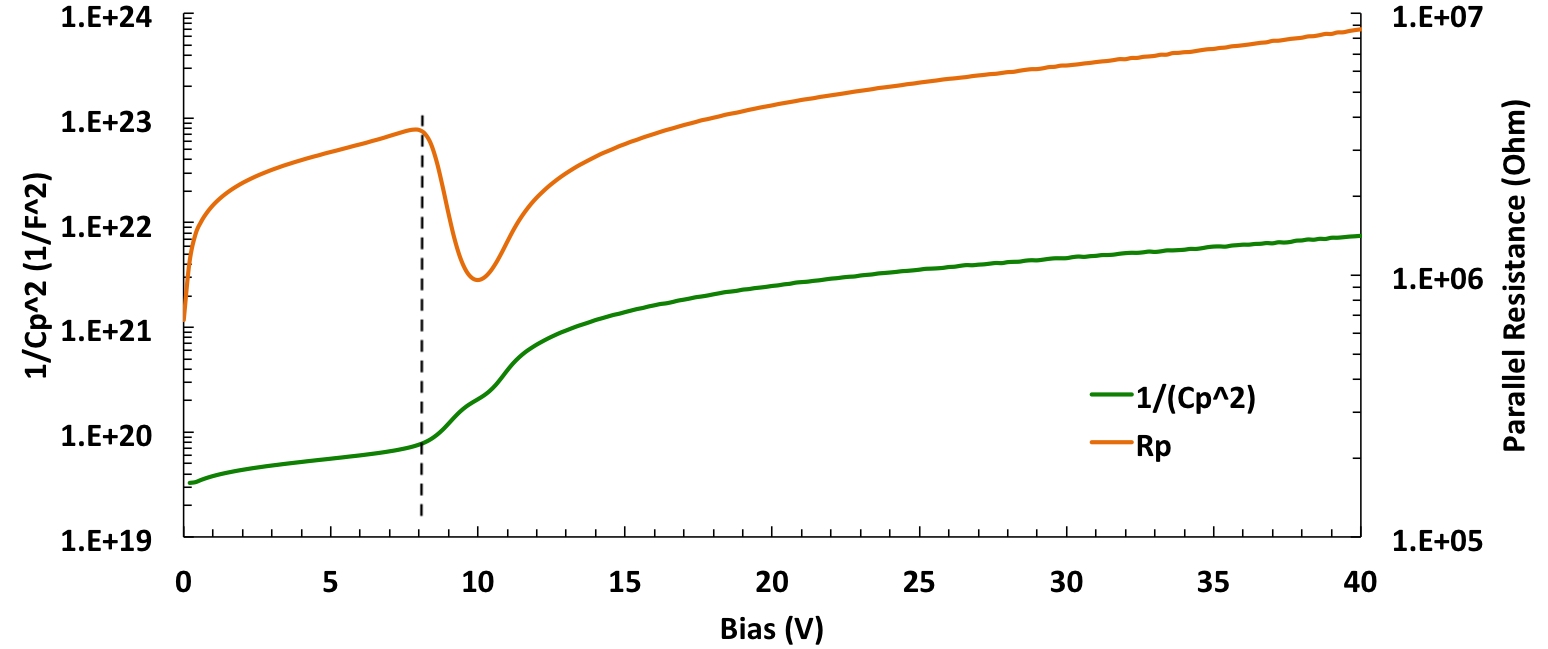}
\caption{This plot shows the  correspondence between $\vglc$ and $\vglr$  for a sensor from W1 irradiated to 3$\nq[15]$. }
\label{fig:Rp}
\end{center}
\end{figure}

The measurements were taken with the Keysight B1505A parameter analyzer using as the model of the silicon detector a  $C_p - R_p$ circuit.
The 1/C$^2$-V curves were obtained at room temperature with a probing frequency of 1 kHz. The value of the frequency was varied between 1 and 3 kHz  finding no  dependence of the results on the operating frequency. Analyzing  how $R_p$ changes with bias, we noticed that in coincidence with $V^{C}_{GL}$  the $R_p$ curve presents a sharp decrease, allowing for an easy identification of the exact voltage of the gain layer depletion.  We indicate $\vgl$ measured with the $R_P$-V curves with the symbol $\vglr$.  The  correspondence between $\vglc$ and $\vglr$ is shown in Figure ~\ref{fig:Rp} for a sensor from W1 irradiated to 3$\nq[15]$.

In the following analysis, the gain layer depletion voltage has been determined using a combination of  the $\vglc$ and $\vglr$ values: at low fluences both $\vglc$ and $\vglr$  are easily identifiable, while for fluences above 1$\nq[15]$ the position of $\vglr$ is easier to identify.   The combination of $\vglc$ and $\vglr$  allows determining  $\vgl$ with an uncertainty  of 0.5 V. 

An interesting parameter to understand the acceptor removal mechanism is the spatial extension of the gain layer. Table~\ref{table:width} reports,  in arbitrary unit, the measured FWHM of the gain layer implants for the wafers exposed to irradiation. The implant widths have been extracted from the doping profiles obtained from the 1/C$^2$-V  curves using the relationship:

\begin{equation}
\label{eq:CV}
N(w) =  \frac{2}{q\epsilon A^2 } \frac{1}{d (1/C(V)^2 )/dV}	 \;\; w =  \frac{\epsilon A^2 } {C(V)},
\end{equation}

where $N(w)$ is the doping density at a depth $w$ and $A$ is the diode's area.

\begin{table}[h]
\begin{center}
\begin{tabular}{|c|c|c|c|}
\hline
Wafer $\#$ &	Dopant	& Gain Dose &   Width [a.u.] \\ \hline \hline
\bf{1} &	B LD &	0.98 &   1 \\
\bf{3} &	B &	1.00 &   1.3 \\
	\bf{6} &	B + C	 &1.02 & 1.3    \\
	\bf{8} &	B	 &1.02     & 	1.3 \\
	\bf{14} &	Ga	 &1.04      &  2.0 \\
	\bf{15} &	Ga + C	 &1.04 &  1.7  \\ \hline
\end{tabular}
\caption{Gain layer FWHM of the wafers used in the irradiation campaign}
\label{table:width}
\end{center}
\end{table}

These widths are consistent with the observation reported in~\cite{shimizu} that carbon co-implantation might yield to narrower implant widths. 

\clearpage 
\section{Irradiation campaign}
Table~\ref{table:ir} reports the wafers and the irradiation steps used in the irradiation campaign.  
A set of LGADs was irradiated without bias  with neutrons  in the JSI research reactor of TRIGA type in Ljubljana. The neutron spectrum and flux are well known ~\cite{JSI-Lj} and the fluence is quoted in 1 MeV equivalent neutrons per $cm^2$ ($n_{eq}/cm^2$).  A different set of LGADs was irradiated with protons at the IRRAD CERN irradiation facility~\cite{CERN-Irr}. The IRRAD proton facility is located on the T8 beam-line at the CERN PS East Hall where the primary proton beam with a momentum of 24 GeV/c is extracted from the PS ring. In IRRAD, irradiation experiments are performed using the primary protons, prior reaching the beam dump located downstream of the T8 beam line.   After irradiation, the devices were annealed for 80 min at 60 $^o$C. Afterward, the devices were kept in cold storage at -20 $^o$C.  The table reports the actual number of protons: the fluences in $n_{eq}/cm^2$ can be obtained by multiplying the proton fluences by the NIEL factor (NIEL =  0.6).

\begin{center}
\begin{table}[h]
\begin{tabular}{|c|c|c|c|c|c|c|}
\hline
Wafer $\#$ &	Dopant	& Gain Dose &   n fluence  [$10^{15} n_{eq}/cm^2$] & p fluence   [$10^{15} p/cm^2$] \\ \hline \hline
\bf{1} &	B LD &	0.98 &    0.2, 0.4, 0.8, 1.5, 3.0, 6.0  &  \\
	\bf{3} &	B 	 &1.00 &   & 0.2, 0.9, 3.9  \\
	\bf{6} &	B + C	 &1.02 &   0.2, 0.4, 0.8, 1.5, 3.0, 6.0& 0.9, 3.9  \\
	\bf{8} &	B	 &1.02     &	 0.2, 0.4, 0.8, 1.5, 3.0, 6.0 &  \\
	\bf{14} &	Ga	 &1.04      &   0.2, 0.4, 0.8, 1.5, 3.0, 6.0	 & 0.9, 3.9\\
	\bf{15} &	Ga + C	 &1.04 &  0.2, 0.4, 0.8, 1.5, 3.0, 6.0 & 0.9, 3.9\\ \hline
\end{tabular}
\caption{Wafers and fluences used in the irradiation  campaign.}
\label{table:ir}
\end{table}
\end{center}

\section{Simulation of different initial acceptor removal rate}
\begin{figure}[htb]
\begin{center}
\includegraphics[width=0.9\textwidth]{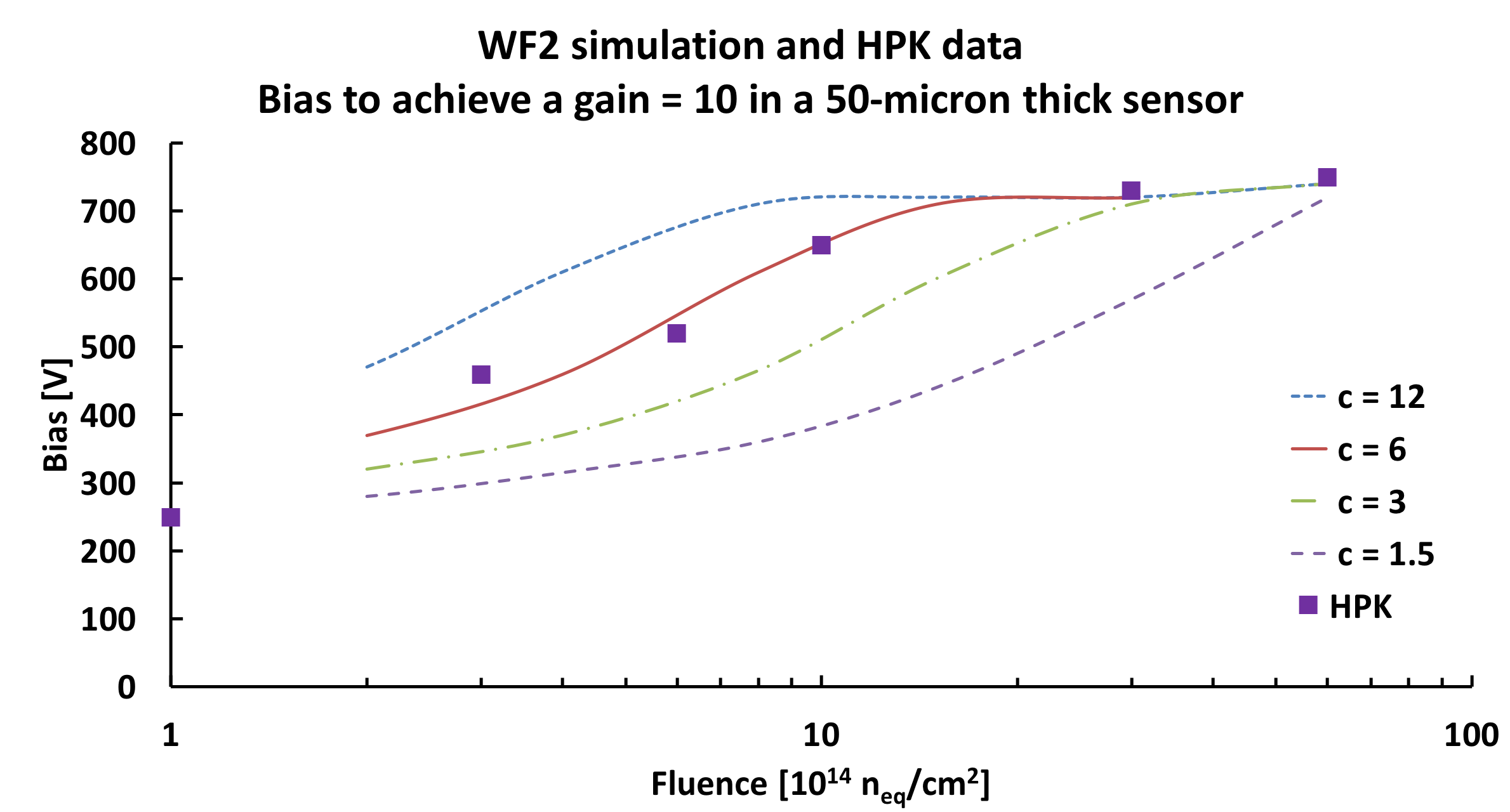}
\caption{Evolution of the bias voltage needed to obtain a constant value of gain, G = 10, as a function of fluence: as the gain layer doping is progressively deactivated by irradiation, the bias voltage is increased  to compensate for the reduction of the electric field generated by the gain layer. The figure shows how a change in the value of the  $c$ exponent (in unit of [$10^{-16} \; cm^2$]) changes this evolution. }
\label{fig:WF2_G10}
\end{center}
\end{figure}

As reported in equation (\ref{eq:ac}), the initial acceptor removal effect is parametrized by the function $c(\NAZ)$. Using the simulation program WF2\footnote{Shareware at http://cern.ch/nicolo}~\cite{WF2},  the effect of larger or smaller values of $c$ on the reduction of the gain has been simulated. Figure~\ref{fig:WF2_G10} reports the bias voltage needed  to keep a constant gain value = 10 as a function of neutron fluence for the situation where the value of $c(\NAZ)$  is twice, a half or a quarter of the presently measured  value of $c  (\NAZ) = 2-3 \cdot 10^{-16}\;cm^{-3})\sim 6 \cdot 10^{-16} \; cm^2$. The simulation has been calculated using the parametrization shown in equation (\ref{eq:ac}), with  $g_{eff}$ = 0.02 $cm^{-1}$ and the $c$ values (in unit of [$10^{-16} \; cm^2$]) shown in the legend. On the plot, the measured points from Hamamatsu LGADs are also reported \cite{Galloway}.

As Figure~\ref{fig:WF2_G10} shows, when the gain layer doping is progressively 
deactivated by irradiation, the bias voltage should be increased  to compensate for the reduction of the electric field generated by the gain layer. Smaller values of $c$ move the need to increase the bias voltage to progressively higher fluences, making LGAD operation more stable.  

\clearpage 
\section{Results}

\begin{figure}[htb]
\begin{center}
\includegraphics[width=1\textwidth]{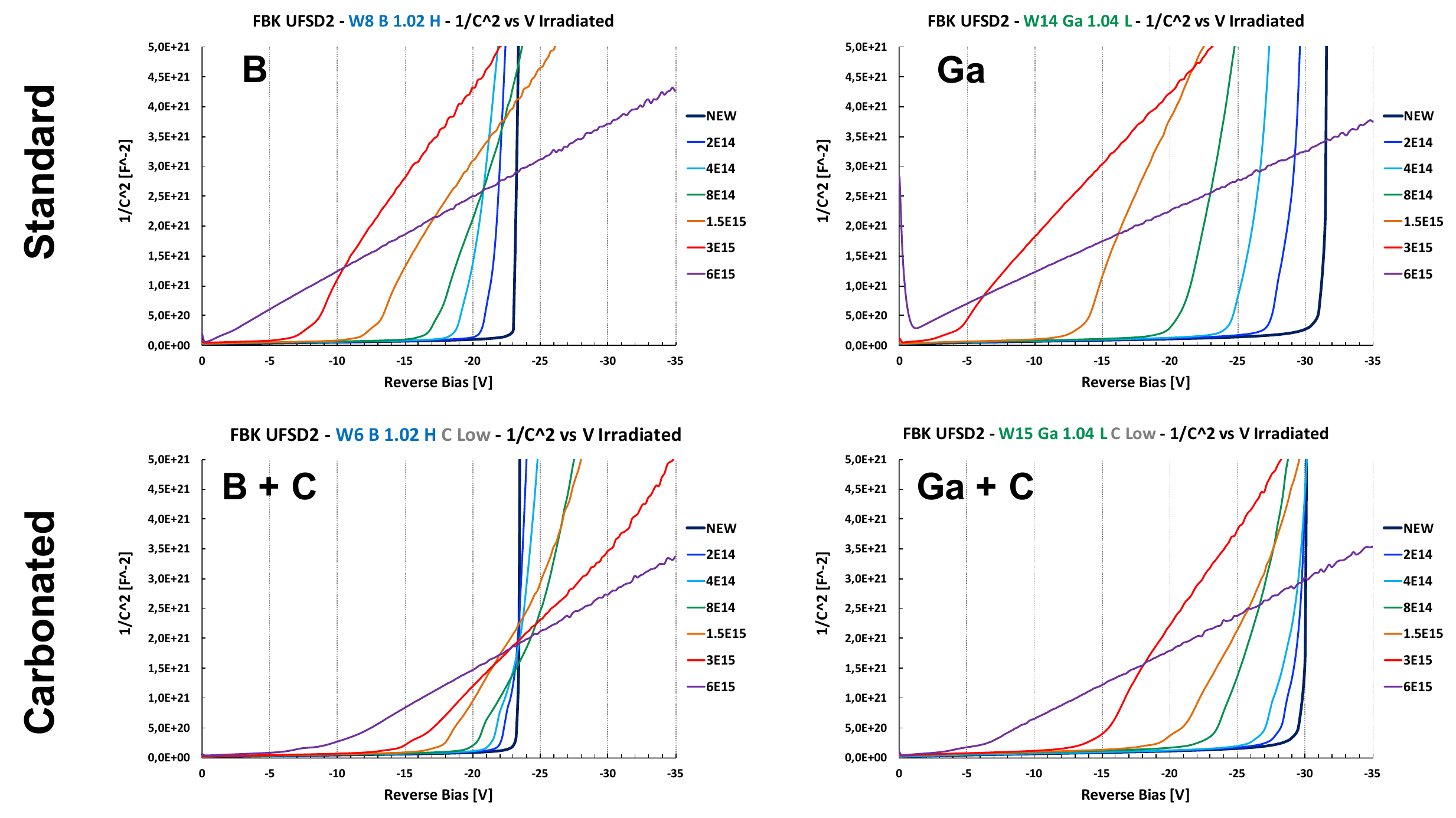}
\caption{Evolution of the 1/C$^2$-V curve  with neutron irradiation for LGAD sensors with different gain layer doping. Irradiation fluence start at $\phi = 2\cdot10^{14} n_{eq}/cm^2$ and double at each step up to $\phi = 6\cdot10^{15} n_{eq}/cm^2$.  Top left: Boron, Top right: Gallium, Bottom left : Boron+Carbon, Bottom right: Gallium+ Carbon}
\label{fig:cv}
\end{center}
\end{figure}

 Figure \ref{fig:cv} shows the evolution of the foot position  ($\vglc$) with increasing neutrons irradiation. The lowest irradiation level is $\phi = 2\cdot10^{14} \;  n_{eq}/cm^2$,  and the fluence increases by a factor of 2 in each of the following curves.

These plots show clearly that the decrease  of $\vglc$ as a function of irradiation for carbonated gain layers is smaller than that of non-carbonated gain layers: for equal fluence, carbonated gain layers retain a higher active doping. Comparing the 4 plots in Figure \ref{fig:cv}, it is evident that the slopes of the 1/C$^2$ curves at equal fluence are similar, indicating, via equation (\ref{eq:CV}), that the doping of the bulk is evolving in the same way for all sensors.

The $c(\NAZ)$  coefficient can be measured by fitting an exponential function to the fraction of still active gain layer as a function of fluence, as shown in equation~(\ref{eq:acnorm}): 

 \begin{equation}
\label{eq:acnorm}
\frac{V_{GL}(\phi)}{V_{GL}(0)}  = \frac{\rho_A(\phi)}{\rho_{A}(0)} = e^{-c(\rho_{A} (0))\phi}.
\end{equation}

\begin{figure}[htb]
\begin{center}
\includegraphics[width=1\textwidth]{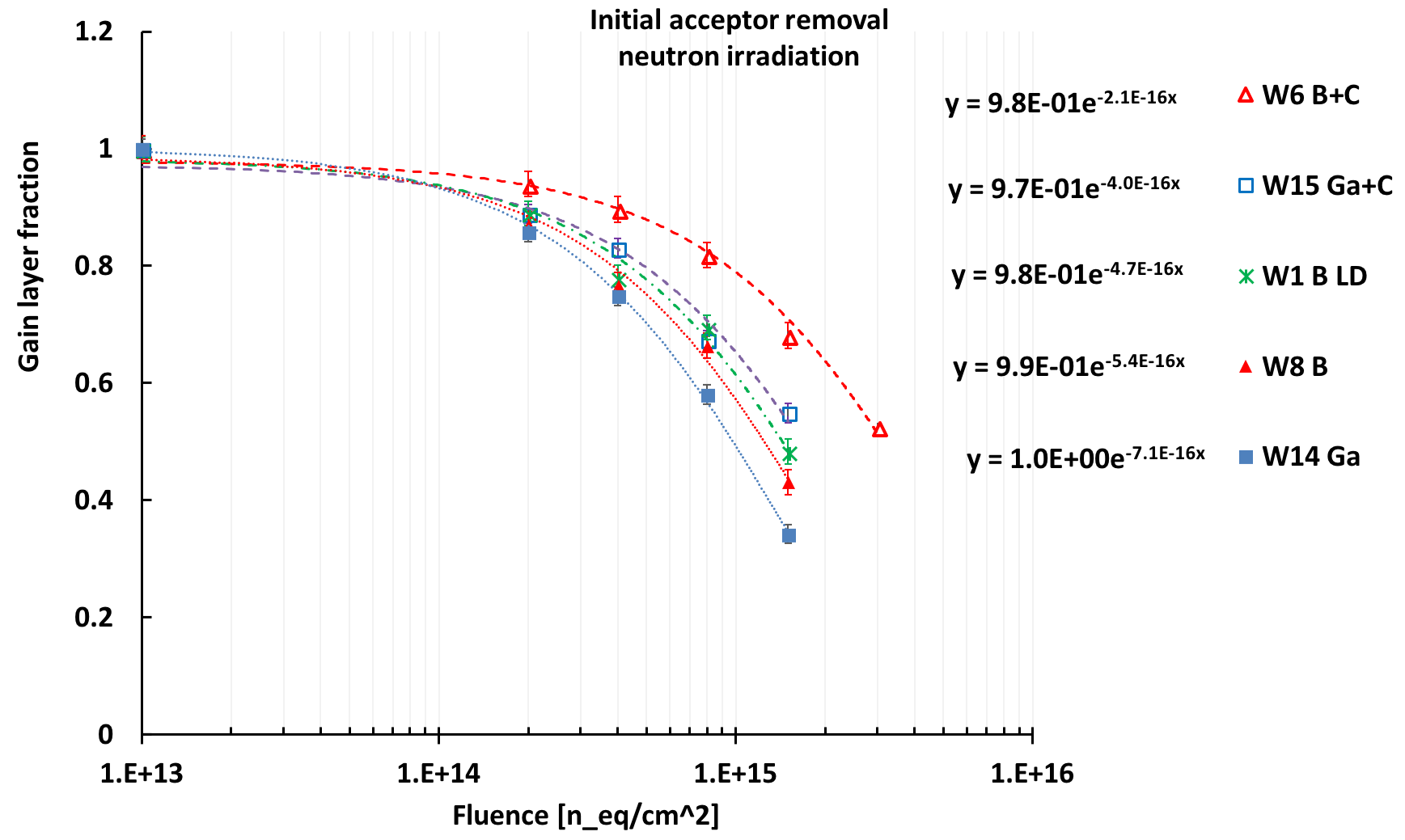}
\caption{Fraction of gain layer still active as a function of neutron irradiation. }
\label{fig:c_n}
\end{center}
\end{figure}

\begin{figure}[htb]
\begin{center}
\includegraphics[width=1\textwidth]{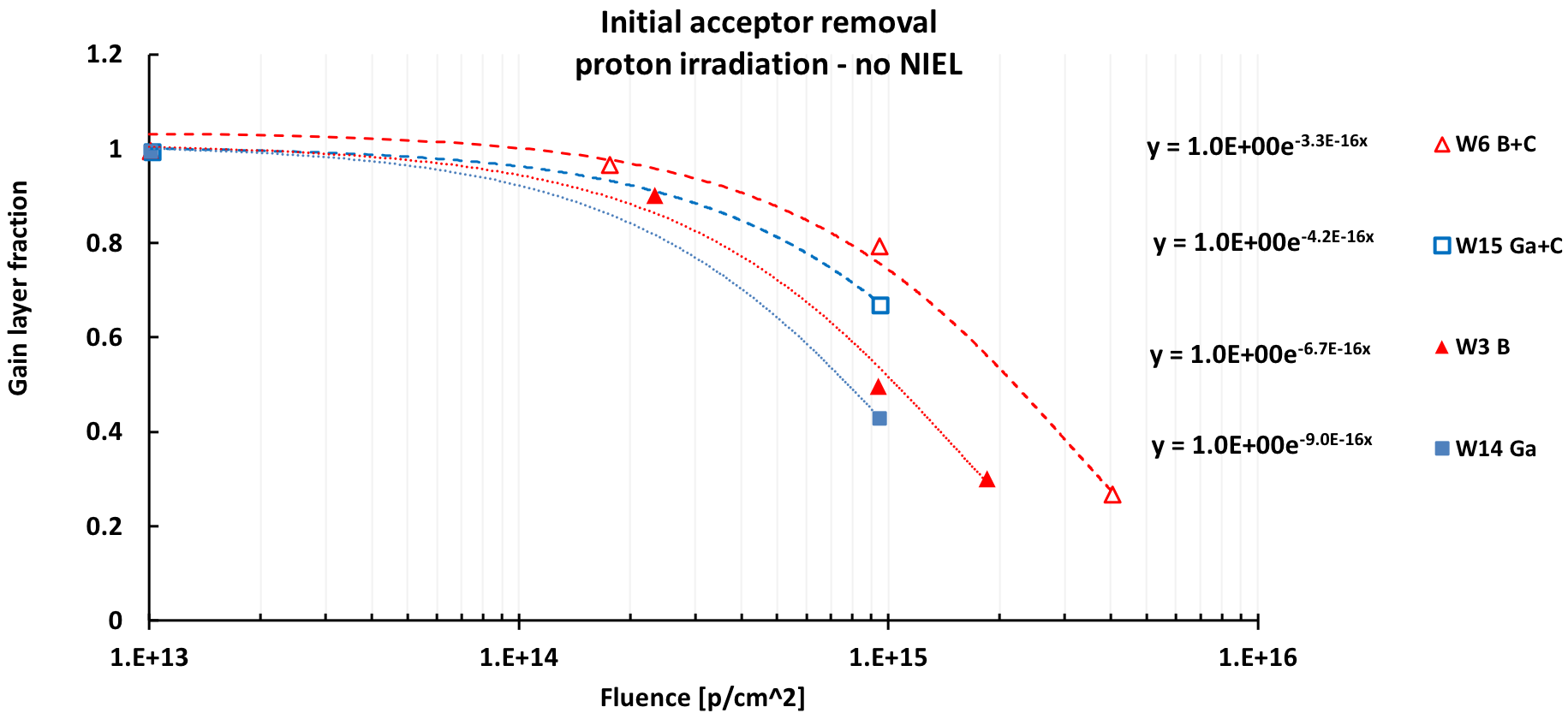}
\caption{Fraction of gain layer still active as a function of proton irradiation.}
\label{fig:c_p}
\end{center}
\end{figure}

The fractions of active gain layer as a function of fluence are shown in  Figure \ref{fig:c_n}  for neutron irradiation and in Figure \ref{fig:c_p} for proton irradiation, together with the exponential fits. 

Table \ref{table:c_coef} reports the compilation of measured values of $c$ for neutron  ($c_n$) and proton ($c_p$) irradiation, and their ratios, ordered in decreasing value. The value of each coefficient has been estimated averaging the measurements of 2 irradiated samples. From the spread of the two measurements, and the uncertainty of the fit,  an error  of $\er$  has been assigned to the determination of $c_n$  while, given the presence of only one measurement per fluence,  the error on $c_p$ has been evaluated to be $\err$. 
\begin{table}[h]
\begin{center}
\begin{tabular}{|c|c|c|c||c|c|}
\hline
Gain Layer &	$c_n$  	& $c_p$ &   $c_n/c_p$  & $c_p $ &   $c_n/c_p$\\
&  $[10^{-16} \; cm^2]$	&  $[10^{-16} \; cm^2]$ &    & $[10^{-16} \; cm^2]$ &   \\ 
& 	& No NIEL &   No NIEL  & NIEL &   NIEL \\ \hline \hline
Ga 		 & 7.1 $\er$ 	&9.  $\err$  & 0.79 $\pm\;0.22$  &15.  $\err$  & 0.47 $\pm\;0.08$\\
B   		&5.4 $\er$ 	&6.5 $\err$ &   0.83 $\pm\;0.29$  &10.8 $\err$ &   0.50 $\pm\;0.11$\\
B LD 	&4.7	 $\er$ 	& &  &&  \\
Ga + C 	&4.0	 $\er$ 	&4.2 $\err$  &  0.95  $\pm\;0.43$ &7.0 $\err$ &   0.57 $\pm\;0.19$ \\ 
B + C 	&2.1	$\er$ 	&3.3 $\err$   & 	0.63  $\pm\;0.66$ &5.5 $\err$ &   0.38 $\pm\;0.54$\\ \hline
\end{tabular}
\caption{Compilation of the initial acceptor removal coefficient for neutrons $c_n$ and protons $c_p$ irradiation for an initial doping density of $\rho(0) \sim 1-2 \cdot 10^{16} $ atoms/cm$^3$. The third column shows the ratio $c_n/c_p$. The error on the $c_n$ has been estimated to be  $\er$ while on $c_p$ is $\err$. The fourth and fifth columns report the $c_p$ values when the NIEL factor has been applied to the proton fluence.}
\label{table:c_coef}
\end{center}
\end{table}

For clarity, Table \ref{table:phi_coef} reports the value of the fluence $\phi_o$ for neutrons and protons. Since the coefficient $\phi_o$ represents the flux needed to remove $63\%$  of the initial acceptor, Table~\ref{table:phi_coef} shows that a carbonated gain layer  can withstand more than twice the radiation of a non-carbonated gain layer.
\begin{table}[h]
\begin{center}
\begin{tabular}{|c|c|c|}
\hline
Gain Layer &	$\phi_o^n  \; [10^{16} \; cm^{-2}]$	& $\phi_o^p \; [10^{16} \; cm^{-2}] $ \\
&  neutrons irrad. & protons irrad.     \\ \hline \hline
Ga 		 & 0.14 $\pm$ 0.02	& 0.11 $\pm$ 0.02 \\
B   		&0.18  $\pm$ 0.03	&0.15  $\pm$ 0.04\\
B LD 	& 0.21 $\pm$	0.05 	&  \\
Ga + C 	&0.25 $\pm$	0.06 	&  0.24 $\pm$ 0.09\\ 
B + C 	&0.48 $\pm$	 0.23	&  0.30 $\pm$ 0.14\\ \hline
\end{tabular}
\caption{Compilation of the initial acceptor removal coefficient $\phi_o$ for neutrons and protons irradiations. As explained in the text, $\phi_o$ represents the flux needed to remove $63\%$  of the initial acceptors.  }
\label{table:phi_coef}
\end{center}
\end{table}

\clearpage

\section{Analysis}

Several results can be extracted from  Table \ref{table:c_coef} :
\begin{itemize}
\item The addition of Carbon improves the radiation resistance: the $c_n, c_p$ coefficients are about a factor of two smaller for B+C  and Ga+C LGADs with respect of those of B or Ga. Since no other condition besides the addition of Carbon was changed, we can determine that the presence of Carbon reduces the value of the coefficient $\kap$. 
\item Considering the real value of proton fluences, the measured $c_p$ and $c_n$ coefficients  are compatible with each other, albeit the $c_p$ values are consistently higher.  This effect indicates that the cross section to remove an acceptor, $\ssi = \kap \cdot N_{Int} \cdot \sigma_{Si}$, is similar for a 1 MeV neutron and a 24 GeV  proton.
\item If the NIEL factor is applied to the protons fluence  (NIEL = 0.6 for 24 GeV/c protons),   the $c_p$ factors are almost twice $c_n$.
\item Narrower and more doped  gain layer implants are less prone to initial acceptor removal:  B LD has a lower $c_n$ coefficient than B. This is consistent with the expectation from the right pane of Figure~\ref{fig:nr} that shows that the relative importance of acceptor removal decreases with increasing initial doping density $\rho_A(0)$.
\item The measured coefficients $c_p, c_n$ for Gallium doping are larger than those for Boron doping. This difference is partly due to the lower Gallium density used in W14 with respect of the Boron density in W3 and W8, however, the difference is larger than what it would be just  due to this effect. This fact might indicate a higher acceptor removal rate of Gallium doping with respect of that of Boron doping. In \cite{KRAMBERGER201853}, a lower  acceptor removal rate of Gallium has been measured with respect of the data reported in this work,  however, the reason might be that the initial Gallium density in \cite{KRAMBERGER201853} was higher than that of this work.

\end{itemize}

\begin{figure}[htb]
\begin{center}
\includegraphics[width=1\textwidth]{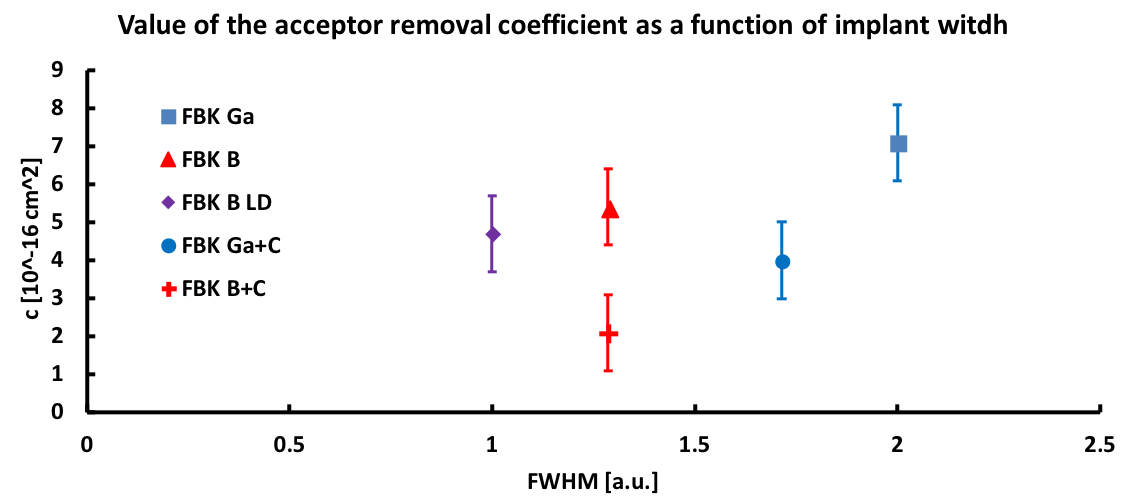}
\caption{Initial acceptor removal coefficient $c_n$ as a function of the gain layer implant  width for carbonated and non-carbonated gain layers: for wider implants the initial acceptor removal mechanism is faster. }
\label{fig:width}
\end{center}
\end{figure}

The gain in LGADs  is required to be 20 - 30: this fact determines  that the total amount of doping in the gain layer is roughly a constant in every LGAD. This given amount of doping can be distributed over narrower or wider implants, varying the doping density: equation~(\ref{eq:ac5}) predicts that in LGADs with wider and less doped implants the initial acceptor removal mechanism is faster. The values of the $c_n$ coefficients as a function of the implant widths reported in Table \ref{table:width}  are shown in  Figure~\ref{fig:width}: the plot clearly shows that in wider implants  the initial acceptor removal mechanism is faster. This effect holds true also for carbonated gain layers.

\vspace{3mm}

\begin{figure}[htb]
\begin{center}
\includegraphics[width=.9\textwidth]{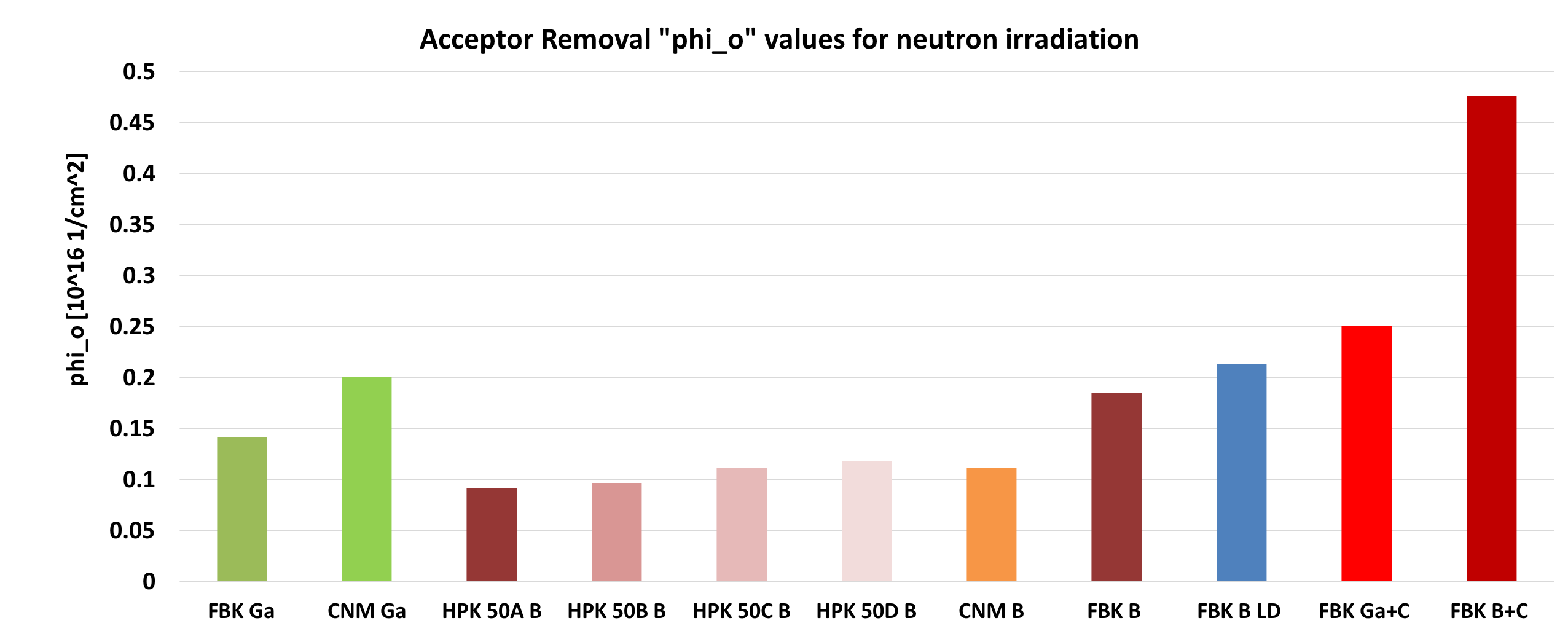}
\caption{Compilation of values of the initial acceptor removal coefficient $\phi_o^n$ for LGADs manufactured by 3 different foundries (HPK, FBK, and CNM) with different gain layer doping compositions.}
\label{fig:coll_c}
\end{center}
\end{figure}

A compilation of values of  $\phi_o$ for neutron irradiation  measured in this work and in ~\cite{1748-0221-10-07-P07006, GK17, HSTD11, TREDI2017, KRAMBERGER201853} is shown in Figure~\ref{fig:coll_c}. All sensors are $\sim$50-micron thick, however, they differ slightly in the doping profile as they don't all have the same gain. 
The plot reports measurements for LGADs manufactured by CNM with a Gallium or a Boron gain layer, 4 different types of  Boron LGADs manufactured by HPK (indicated with the names 50A, 50B, 50C and 50D in order of increasing gain layer doping levels) and several LGADs manufactured by FBK.  The carbonated gain layers have clearly the largest values of $\phi_o$, followed by  B LD: the 1/e fluence for B+C LGADs is almost $0.5\nq[16]$.

\begin{figure}[htb]
\begin{center}
\includegraphics[width=1.\textwidth]{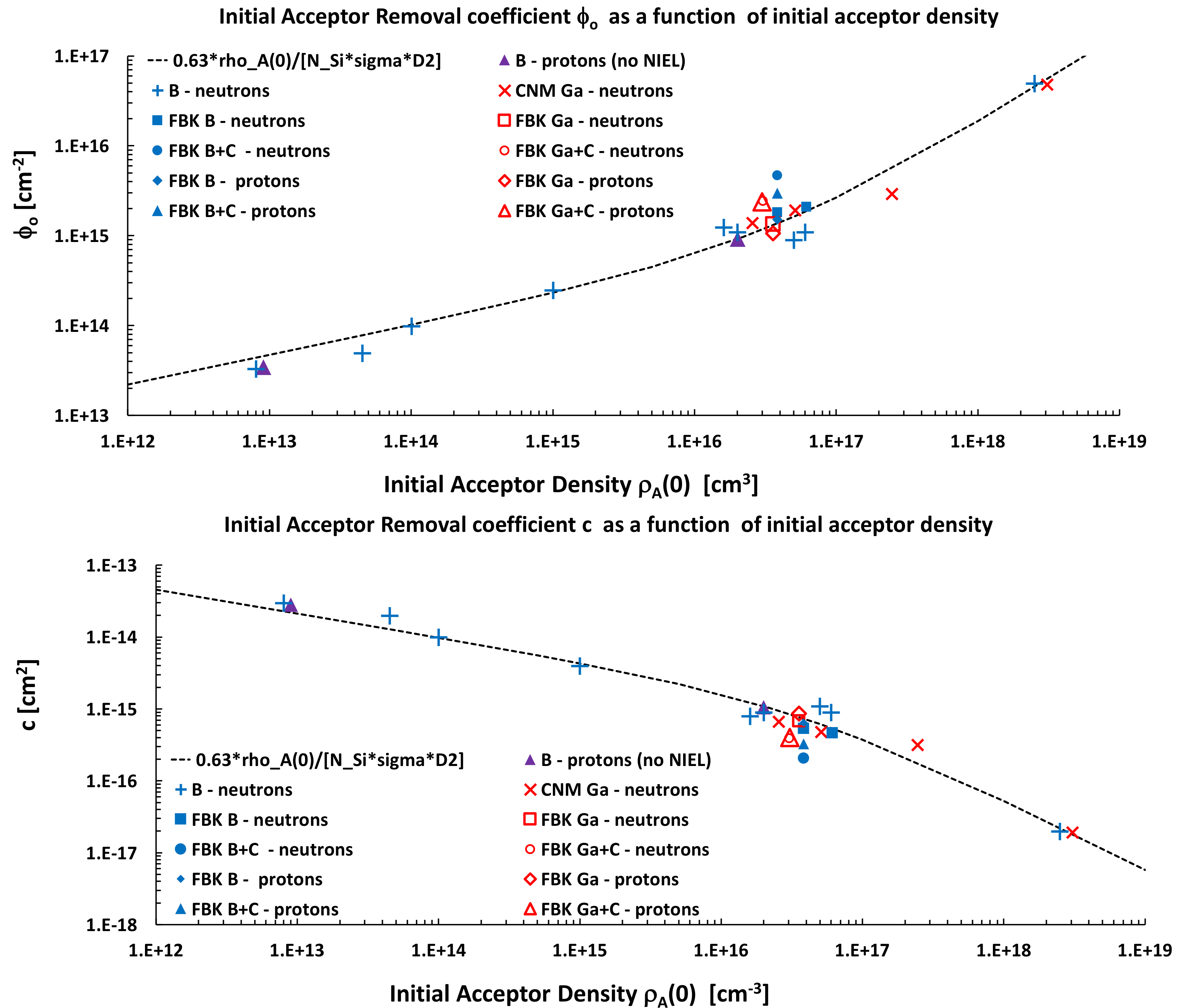}
\caption{Values of the $\phi_o$ and $c$ coefficients from previous measurements and from this analysis.  }
\label{fig:c_FBK}
\end{center}
\end{figure}

 Figure~\ref{fig:c_FBK} updates Figure~\ref{fig:coef} including the results obtained in this analysis: the new points  cluster around $\NAZ \sim 2-6 \cdot 10^{16}$. The value of $\NAZ$ has been obtained by computing the gain layer doping profile using the relationship, shown in equation~(\ref{eq:CV}), between the derivative of the curve $1/C^2 - V$ and the doping at a depth $w$.

\clearpage

\section{Measurement of the gain due to the gain-layer  after a fluence of $\phi = 8\cdot 10^{14}, \;  1.5\cdot 10^{15}$ and $3\cdot 10^{15} \;n_{eq}/cm^2$.}
Using a collimated picosecond laser system with a light spot diameter of $\sim$ 20 microns and a wavelength of 1064 nm,  the gains of B, B LD, B+C, Ga and Ga+C  LGADs were measured as a function of bias voltage for 3 neutron irradiation levels: $\phi = 8\cdot 10^{14}, \;  1.5\cdot 10^{15}$ and $3\cdot 10^{15} \;n/cm^2$. The value of the gain was obtained as the ratio of the signal areas obtained in an LGAD and in a PiN diode irradiated to the same fluence. 

\begin{figure}[htb]
\begin{center}
\includegraphics[width=1\textwidth]{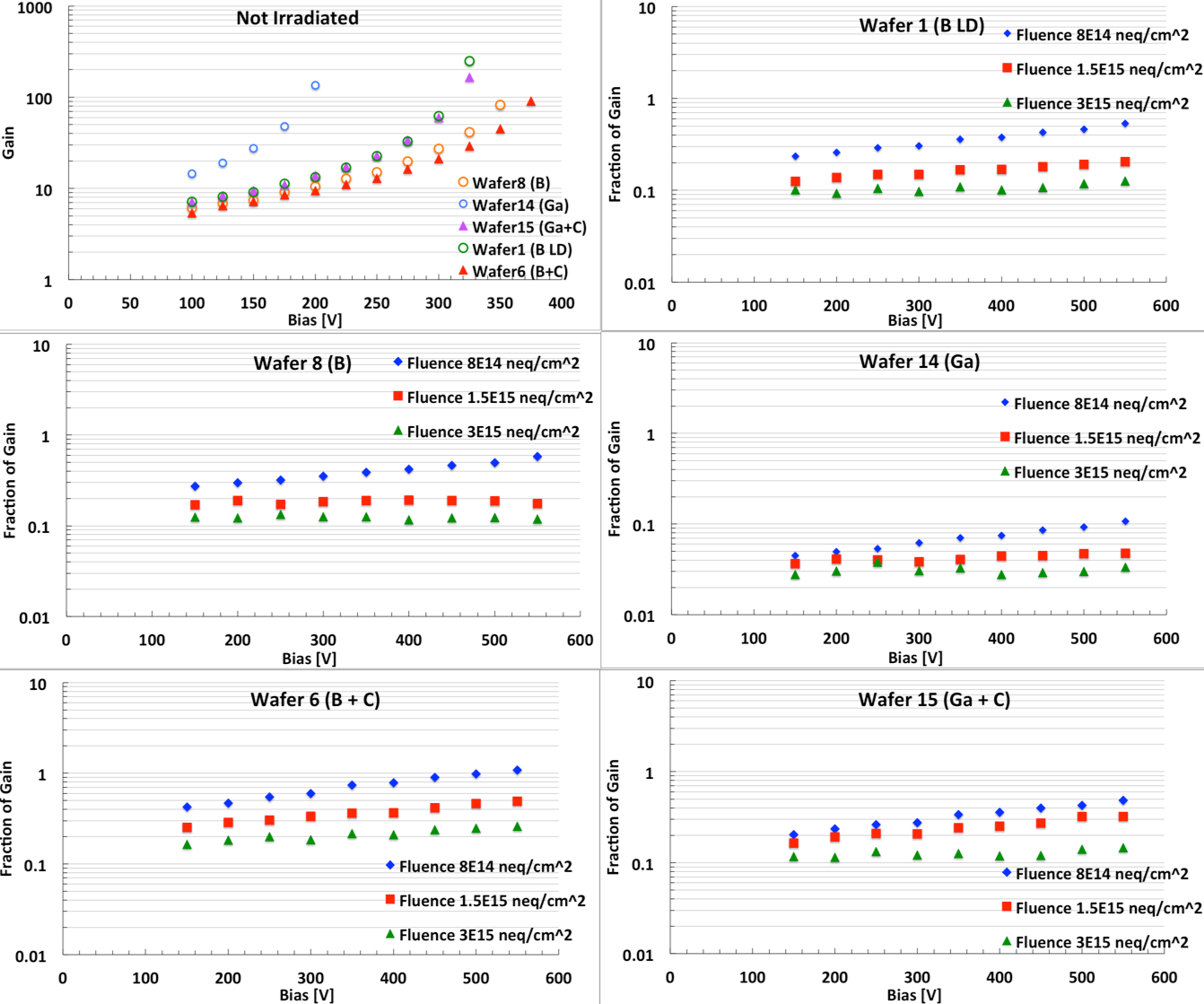}
\caption{Top left plot: gain curves before irradiation. Following 5 plots: for each gain layer type, the plot shows the fraction of gain at 3 fluences normalized to each respective gain at Bias = 150V. }
\label{fig:gain}
\end{center}
\end{figure}

The results are shown in Figure~\ref{fig:gain}: the top left plot shows the gain curves before irradiation, while the following 5 plots show the gain  normalized to the  respective unirradiated gain at Bias = 150V. As expected, B+C is the most radiation resistant LGAD: after a fluence of $8\nq[14]$ the gain layer still generate at bias = 500V  the same gain as it had when not irradiated at bias = 150V.  Likewise, Ga is the weakest retaining at 500V only 10\% of the initial gain.

Confirming the results on the values of the $c_n$  coefficient, carbonated gain layers (B+C and Ga+C) show higher gain values   than those without Carbon for the same fluence level.  Likewise, B LD maintains higher gain values than B; at $\phi = 3\cdot 10^{15} \;n/cm^2$  only B+C gain layer is still active. It is possible that by optimizing the Carbon dose  this effect can be further enhanced. 

\section{Conclusions and outlook}
 50-micron thick LGADs manufactured  by FBK with 5 different types of gain layer doping (B, B+C, Ga, Ga+C and B  LD) have been irradiated with neutrons and protons.  The results show that (i) carbonated gain layer are at least a factor of two more radiation resistant than the equivalent non-carbonated gain layer, (ii) Gallium doping is less radiation resistant than Boron doping, (iii) narrower gain layer implants are more radiation resistant than wider implants,  (iv) considering the true fluence value, protons with 24 GeV/c momentum are similarly   harmful than 1 MeV neutrons  with respect of the initial acceptor removal mechanism, and that (v) if the fluence of protons with 24 GeV/c momentum is converted using the NIEL factor to 1 MeV equivalent neutrons, proton irradiation is much more harmful than that from 1 MeV neutrons .

\vspace{2mm}

Carbonated gain layer holds the possibility of designing silicon sensors with gain with enhanced radiation resistance. We plan to further investigate the property of carbonated gain layer by producing gain layers with several carbon doses, to optimize the radiation resistance of the LGAD design.  We are confident that these findings, albeit obtained for LGAD sensors, can be successfully implemented in other silicon sensors with gain such as SiPM and APD.

\section{Acknowledgments}
We acknowledge the fundamental contributions coming from the discussions, and active collaboration of the RD50 colleagues. We recognize the key contributions of the irradiation facilities at Ljubjiana and IRRAD at CERN.
Part of this work has been financed by the European Union's Horizon 2020 Research and Innovation funding program, under Grant Agreement no. 654168 (AIDA-2020) and Grant Agreement no. 669529 (ERC UFSD669529), and by the Italian Ministero degli Affari Esteri and INFN Gruppo V

\section*{References}

\bibliography{acc_removal_arxiv_II.bbl}

\end{document}